\documentclass[fleqn,useAMS,usenatbib]{mnras}
\usepackage{graphicx}
\usepackage[english]{babel}
\usepackage{amsmath}
\usepackage{amssymb}
\usepackage{natbib}
\usepackage{wasysym}
\usepackage{newtxtext,newtxmath}
\usepackage[T1]{fontenc} 

\DeclareMathOperator{\rp}{r-process}

\title[r-process enrichment with neutron star kicks]{The impact of natal kicks on galactic r-process enrichment by neutron star mergers}
\author[F.~van~de~Voort et al.]{Freeke van~de~Voort,$^{1}$\thanks{E-mail: freeke@astro.cf.ac.uk} R\"udiger Pakmor,$^{2}$ Rebekka Bieri$^{2}$ and Robert~J.~J.~Grand$^{3,4}$ \\
$^1$Cardiff Hub for for Astrophysics Research and Technology, School of Physics and Astronomy, Cardiff University, Queen's Buildings, The Parade, Cardiff \\ CF24 3AA, UK \\
$^2$Max Planck Institute for Astrophysics, Karl-Schwarzschild-Stra{\ss}e 1, 85748, Garching, Germany \\
$^3$Instituto de Astrof\'isica de Canarias, Calle Vía L\'actea s/n, E-38205 La Laguna, Tenerife, Spain\\
$^4$Departamento de Astrof\'isica, Universidad de La Laguna, Av. del Astrof\'isico Francisco S\'anchez s/n, E-38206, La Laguna, Tenerife, Spain \\
}

\begin{document}

\date{Accepted 2022 March 10. Received 2022 March 10; in original form 2021 October 21}

\pagerange{\pageref{firstpage}--\pageref{lastpage}} \pubyear{2021}

\maketitle

\label{firstpage}

\begin{abstract}
We study galactic enrichment with rapid neutron capture (r-process) elements in cosmological, magnetohydrodynamical simulations of a Milky Way-mass galaxy. We include a variety of enrichment models, based on either neutron star mergers or a rare class of core-collapse supernova as sole r-process sources. For the first time in cosmological simulations, we implement neutron star natal kicks on-the-fly to study their impact. With kicks, neutron star mergers are more likely to occur outside the galaxy disc, but how far the binaries travel before merging also depends on the kick velocity distribution and shape of the delay time distribution for neutron star mergers. In our fiducial model, the median r-process abundance ratio is somewhat lower and the trend with metallicity is slightly steeper when kicks are included. In a model `optimized' to better match observations, with a higher rate of early neutron star mergers, the median r-process abundances are fairly unaffected by kicks. In both models, the scatter in r-process abundances is much larger with natal kicks, especially at low metallicity, giving rise to more r-process enhanced stars. We experimented with a range of kick velocities and find that with lower velocities, the scatter is reduced, but still larger than without natal kicks. We discuss the possibility that the observed scatter in r-process abundances is predominantly caused by natal kicks removing the r-process sources far from their birth sites, making enrichment more inhomogeneous, rather than the usual interpretation that the scatter is set by the rarity of its production source.  
\end{abstract}


\section{Introduction} \label{sec:intro}
 
Neutron stars in the Milky Way have, on average, velocities in the range of hundreds of km s$^{-1}$ \citep[e.g.][]{Hobbs2005, Faucher2006}. Some have even been observed with velocities exceeding 1000 km s$^{-1}$. This is not the case for their progenitor population, i.e.\ massive stars. Instead, these high velocities are imparted on the neutron stars when they are born, which is why these are referred to as natal kicks. A likely explanation for these kicks are asymmetries in the supernova explosions that expel the outer material of the progenitor stars, causing the resulting compact object to be born with a significant kick velocity \citep[e.g.][]{Scheck2006, Janka2013}. This means that neutron stars, including binary neutron stars, are more likely to be found outside the disc of the galaxy than massive main sequence stars. 

Whether or not a binary star system survives the natal kick imparted by its supernovae depends on the exact binary configuration and the dynamics of the supernova explosions \citep[e.g.][]{Bray2016}. The binary system is more likely to remain bound if the natal kick it experiences is weaker. The resulting velocity distribution of binary neutron stars may therefore be different, with a lower average velocity, than that of single neutron stars. The handful of binary neutron stars known in the Galaxy have estimated velocities ranging from $\approx0-500$~km s$^{-1}$ \citep{Wong2010}. Short gamma-ray bursts, whose progenitors are thought to be neutron star mergers, show substantial offsets from their host galaxies. Using these offsets, kick velocites of $\approx20-140$~km s$^{-1}$ have been derived, consistent with the Galactic estimate \citep{Fong2013}. For both single and double neutron stars, the sample sizes are small and the uncertainties in kick velocities are therefore large.

Neutron star mergers are promising sites of production of rapid neutron capture (r-process) elements. The distribution of elements produced in such explosions is reasonably consistent with observations of the Sun and other stars, even for the heaviest r-process elements \citep[e.g.][]{Lattimer1977, Freiburghaus1999, Bauswein2013, Ji2016b}. Furthermore, their rarity can potentially explain the observed scatter in r-process abundances at low metallicity as probed by europium, an element that is almost exclusively produced by the r-process \citep[e.g.][]{Burris2000}. The first detection of gravitational waves from a neutron star merger led to the detection of an electromagnetic counterpart, called a kilonova \citep[e.g.][]{Abbott2017, Coulter2017}. The evolution of the light curves showed rapidly declining blue emission and long-lived red emission, which is consistent with the kilonova being powered by the radioactive decay of high-opacity r-process elements and difficult to explain otherwise \citep[e.g.][]{Drout2017, Kasen2017, Pian2017}. This lends further credence to the theory that r-process elements are predominantly produced in kilonovae. 

However, idealized chemical evolution models and cosmological simulations using neutron stars as the only source for r-process elements struggle to reproduce observations of stellar europium abundances in the Milky Way in detail \citep[e.g.][]{Argast2004, Matteucci2014, Wehmeyer2015, Cote2019, Haynes2019, Voort2020}. A possible solution is that there is a second source for r-process elements, which could even dominate over r-process element production by neutron star mergers. One candidate for such a source is a special, rare type of core-collapse supernova, such as magneto-rotational supernovae or collapsars \citep[e.g.][]{MacFadyen1999, Cameron2003, Fujimoto2008, Winteler2012, Mosta2018, Halevi2018, Siegel2019, Cowan2021}. Other potential r-process sources include, for example, magnetized neutrino- driven winds from proto-neutron stars \citep{Thompson2018} and common envelope jet supernovae \citep{Grichener2019}. Such a source should be rare enough that it produces the large observed scatter in europium at low metallicity, but not so rare that a large fraction of metal-poor stars are born without r-process elements. 

It is also possible that the current generation of neutron star merger models are missing a key ingredient and that including this ingredient in our models would provide a better match with observations. The potential importance of natal kicks for r-process enrichment has recently been highlighted by \citet{Banerjee2020}, but has not yet been explored in cosmological simulations. Galaxy formation is complex and governed by many processes, such as large-scale outflows, gas accretion and reaccretion, (metal) mixing, galaxy mergers, and satellite stripping. All of these processes interact with one another, adding even more complexity. Cosmological, (magneto)hydrodynamical simulations, which base their initial conditions on fluctuations observed in the cosmic microwave background radiation, include all of these processes and their interactions. They can be thus used to follow inhomogeneous chemical evolution on-the-fly over the full history of the Universe. Previously, we used such simulations to follow r-process enrichment from neutron star mergers embedded in the existing stellar population \citep{Voort2020}. We have now implemented neutron star natal kicks to study their impact on the resulting stellar r-process abundances. 

In this work, we report on the results from our cosmological simulations and focus on r-process enrichment by neutron star mergers, but also contrast this with models for enrichment via rare core-collapse supernovae. For the first time, we study the effect of neutron star natal kicks, varying the uncertain parameters of the neutron star delay time distribution and the natal kick velocity distribution. We also include a model for extremely rare core-collapse supernovae in our high-resolution simulation, which we previously had only simulated at medium resolution \citep{Voort2020}. In Section~\ref{sec:sim} we briefly discuss our cosmological set-up and our galaxy formation model. The way we implemented natal kicks and neutron star mergers is described in Section~\ref{sec:NS} and rare core-collapse supernovae in Section~\ref{sec:SN}. Our results are shown in Section~\ref{sec:results}, where we discuss our fiducial model (Section~\ref{sec:kicks}), the impact of varying our model parameters (Section~\ref{sec:var}), and a model with `optimized' parameters (Section~\ref{sec:optimal}), which we compare to observations (Section~\ref{sec:obs}) and use to test very low kick velocities  (Section~\ref{sec:vel}). We summarize our results and discuss further implications in Section~\ref{sec:concl}.

\section{Method} \label{sec:sim}

This work is based on a simulation from the Auriga project \citep{Grand2017}. Auriga consists of a large number of zoom-in magnetohydrodynamical, cosmological simulations of relatively isolated Milky Way-mass galaxies and their environments, simulated at high resolution to about 1~Mpc from the central galaxy within a lower resolution (100~Mpc)$^3$ volume. The simulated galaxies are disc-dominated and have stellar masses, galaxy sizes, rotation curves, star formation rates, and metallicities in agreement with observations. For this work, we resimulate one of the Auriga galaxies (called `halo~6') at high resolution. We also carried out medium-resolution and low-resolution simulations, again of halo~6, in order to test for resolution effects and additional model variations. The galaxy in our chosen halo has properties that are reasonably similar to the Milky Way. Additionally, it has a relatively compact zoom-in region, which makes it one of the most efficient Auriga haloes in terms of computational time. The simulations were carried out with the quasi-Lagrangian moving-mesh code \textsc{arepo} \citep{Springel2010}.

Our main results focus on a new high-resolution zoom-in simulation. Its target cell resolution is $m_\mathrm{cell}=6.7\times10^3$~M$_{\astrosun}$ and dark matter particle mass is $m_\mathrm{DM}=3.6\times10^4$~M$_{\astrosun}$. We furthermore ran an additional simulation with different neutron star merger parameters, results from which better match currently available observational data, which is shown in Section~\ref{sec:optimal}. This simulation has 8 times lower mass resolution than our fiducial simulation, which is referred to as medium resolution ($m_\mathrm{cell}=5.4\times10^4$~M$_{\astrosun}$ and $m_\mathrm{DM}=2.9\times10^5$~M$_{\astrosun}$). We assumed a $\Lambda$CDM cosmology with parameters $\Omega_\mathrm{m}=1-\Omega_\Lambda=0.307$, $b=0.048$, $h=0.6777$, $\sigma_8=0.8288$, and $n=0.9611$ \citep{PlanckXVI2014}. 

\begin{figure}
\center
\includegraphics[scale=1.0]{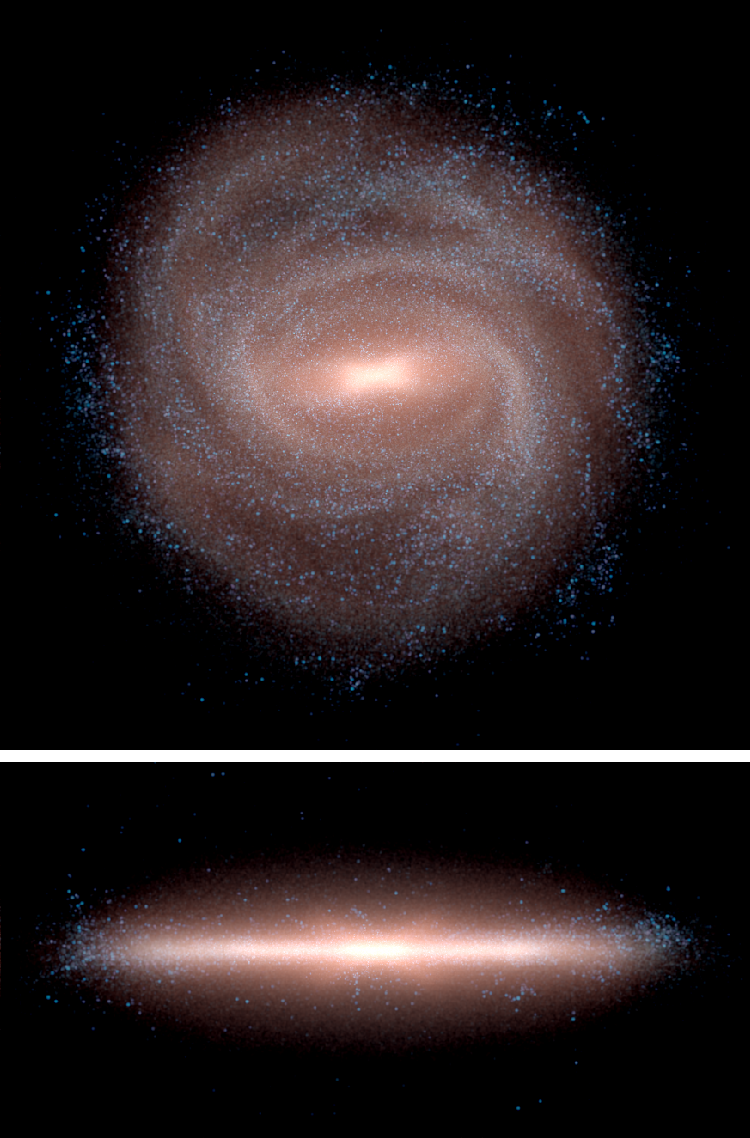}
\caption {\label{fig:img} Face-on and edge-on images of the stars in K-, B-, and U-band (shown in red, green, and blue, respectively) at $z=0$. The images are 50 kpc $\times$ 50 kpc (top panel) and 50 kpc $\times$ 25 kpc (bottom panel). The galaxy exhibits a bar with predominantly older (redder) stars and an extended star-forming (bluer) disc. Its total stellar mass is $7\times10^{10}$~M$_{\astrosun}$ and its present-day SFR is about 3~M$_{\astrosun}$~yr$^{-1}$.} 
\vspace{-3mm}
\end{figure}

Gas becomes star-forming above a density threshold of $n_\mathrm{H}^\star=0.11$~cm$^{-3}$ and forms stars stochastically \citep{Springel2003}. This dense gas is put on an effective equation of state, because we cannot resolve the multi-phase nature of the interstellar medium (ISM). The ISM in our simulations is therefore smoother than it is in reality. At $z=0$, the total stellar mass contained within 30~kpc is $7.1\times10^{10}$~M$_{\astrosun}$. The total stellar mass within the virial radius, $R_\mathrm{vir}$, is $7.8\times10^{10}$~M$_{\astrosun}$, where $R_\mathrm{vir}=214$~kpc and defined as the radius within which the average density is equal to 200 times the critical density of the Universe. The total halo mass within $R_\mathrm{vir}$ is $1.0\times10^{12}$~M$_{\astrosun}$. The low redshift star formation rate (SFR) is $3.0$~M$_{\astrosun}$~yr$^{-1}$ ($3.4$~M$_{\astrosun}$~yr$^{-1}$), averaged over the last Gyr (100~Myr). The full star formation history is shown in \citet{Grand2021} where this simulation is referred to as `Level 3'. A 3-colour face-on and edge-on image of the galaxy is shown in Figure~\ref{fig:img} in which older stars appear redder and younger stars appear bluer. The galaxy is disc-dominated with a central bar and boxy/peanut bulge \citep[see e.g.][]{Fragkoudi2020}. 

A star particle in the simulation represents a single stellar population with a specific age and metallicity with an initial mass function from \citet{Chabrier2003}. Stars with masses above 8~$M_{\astrosun}$ explode as core-collapse supernovae. Mass-loss and metal return are calculated for core-collapse and Type Ia supernovae and for asymptotic giant branch (AGB) stars at each simulation time-step using the yields from \citet{Karakas2010} for AGB stars, from \citet{Portinari1998} for core-collapse supernovae, and from \citet{Thielemann2003} and \citet{Travaglio2004} for Type Ia supernovae. The released mass and metals are injected into the single host cell of the star particle. This single cell injection differs from the original Auriga suite of simulations, where metals were injected into 64 neighbouring cells. This change was found to slightly increase the $2\sigma$ scatter of the stellar abundances, but to have little effect on the average or $1\sigma$ scatter \citep{Voort2020}. It thus does not impact our results strongly. The galaxy formation model includes stellar and active galactic nucleus (AGN) feedback, which results in large-scale outflows and limits the growth of the galaxy \citep{Grand2017}. These outflows carry metals out of the ISM and into the halo and therefore have a large impact on the chemical evolution of the galaxy. 

The elements produced by standard and Type Ia supernovae are released at each time step based on the age of the stellar particle, which represents a single stellar population, resulting in a smooth enrichment of the host cell. Although this choice may reduce the scatter in the elemental abundances slightly, the production sites of these elements are ubiquitous and we therefore do not expect our results to be strongly affected. We explicitely follow the elements H, He, C, N, O, Ne, Mg, Si, and Fe. These elements affect the cooling rate of the gas and thus have a dynamical effect on the evolution of the galaxy.

The additional channels that represent r-process elements, whilst also being traced on-the-fly with the same time resolution as other components in the simulation, are implemented as passive tracers that do not affect the rest of the simulation. Their presence is thus inconsequential to the evolution of the galaxy -- a good approximation, because these elements are very rare and their effect is therefore negligible. We choose not to use yield tables, with their inherent large uncertainties, but rather renormalize each r-process enrichment model in post-processing (see below for more details). Although the r-process yields may depend on metallicity, this is highly uncertain and we therefore do not include such a dependence. Each event releases the same amount of r-process elements. In contrast to the more common elements, r-process enrichment is implemented stochastically. The produced elements are released only when an r-process event takes place, rather than smoothly at every time step. Our stochastic r-process sources do not produce any of the more common elements in our simulation. Such a co-production of other elements is more important for rare core-collapse supernovae than for neutron star mergers, though likely still negligible due to the efficient mixing of pristine and enriched gas in our simulations \citep{Voort2020}.

Abundance ratios of our star particles are normalized by the Solar abundances from \citet{Asplund2009} and defined as
\begin{equation}
  [\mathrm{A/B}] = \mathrm{log}_{10}\dfrac{N_\mathrm{A}}{N_\mathrm{B}} - \mathrm{log}_{10}\dfrac{N_\mathrm{A}}{N_\mathrm{B}}_{\astrosun}
\end{equation}
where $A$ and $B$ are two different elements and $N_\mathrm{A}$ and $N_\mathrm{B}$ are their number densities. We use magnesium as a metallicity indicator in this work. Because the magnesium yields used in our simulations are known to be low \citep{Portinari1998}, we multiply them by a factor of 2.5 in post-processing to correct for this \citep[see also][]{Voort2020}. Magnesium is not a dominant coolant in the circumgalactic medium. Therefore, such an increase in the magnesium abundances would not affect the dynamics of our simulations. We choose magnesium and not iron as a metallicity tracer, because magnesium is primarily produced in core-collapse supernovae, whereas iron is produced both in core-collapse and Type~Ia supernovae. Using magnesium is therefore easier to interpret as it does not depend on the evolving ratio of Type~Ia supernovae to core-collapse supernovae. 

Furthermore, because the r-process elements are passive tracers and do not have any dynamical effect on the simulations, we can change the r-process yield in post-processing without loss of generality. Given that the r-process yields are very uncertain, this approach allows us to focus on the trends with metallicity and the scatter in abundance ratios. We choose to normalize the resulting r-process abundance ratios for each of our models separately by setting the median $\mathrm{[\rp/Mg]}$ to 0 at $\mathrm{[Mg/H]}=0$. This is the reason why our models have different europium yields as listed in the last column of Table~\ref{tab:models}.

\subsection{Modeling neutron star natal kicks and mergers} \label{sec:NS}

\begin{table*}
\begin{center}                                                                                                                                        
\caption{\label{tab:models} \small Parameters of r-process enrichment models for those based on neutron star mergers (first 8 rows). Listed in columns 2--5 are the number of neutron star mergers per $M_{\astrosun}$ of stars, the minimum delay time for neutron star mergers in a simple stellar population, the delay time distribution (DTD) power-law exponent, and the median kick velocity and its dispersion. The final two columns show the resulting r-process event rate at $z=0$ (averaged over the last Gyr) and the europium yield ($y_\mathrm{Eu}$) per r-process event (calculated by normalizing the median $\mathrm{[Eu/Mg]}$ to zero at $\mathrm{[Mg/H]}=0$) for the 8 neutron star models and the 2 rare core-collapse supernova models. To obtain the total yield of r-process material of elements with $A>95$, $y_\mathrm{Eu}$ should be multiplied by a factor 215.}  
\begin{tabular}[t]{llrrrrr}
\hline \\[-3mm]                                                                                                                                       
model name   & $A$                                   & $t_\mathrm{min}$              & $\gamma$           & $v_\mathrm{kick}$  & $R_\mathrm{\rp}(z=0)$ & $y_\mathrm{Eu}$ \\
                      & (M$_{\astrosun}^{-1}$)                  & (Myr)                                  &                           & (km s$^{-1}$)      & (yr$^{-1}$)             & (M$_{\astrosun}$)\\
\hline \\[-4mm]                                                                                                                                       
  fiducial kick     & $3\times10^{-6}$                & $30$               & $-1.0$               & $200\pm500$                & $1.2\times10^{-4}$       & $9.7\times10^{-5}$ \\
  no kick     & $3\times10^{-6}$                & $30$               & $-1.0$               & $\mathbf{0}\pm\mathbf{0}$  & $1.0\times10^{-4}$       & $6.5\times10^{-5}$ \\
  small kick     & $3\times10^{-6}$                & $30$               & $-1.0$               & $\mathbf{0}\pm500$        & $1.2\times10^{-4}$       & $8.7\times10^{-5}$ \\
  large kick     & $3\times10^{-6}$                & $30$               & $-1.0$               & $\mathbf{500}\pm500$     & $1.3\times10^{-4}$       & $11.9\times10^{-5}$ \\
  narrow range    & $3\times10^{-6}$                & $30$               & $-1.0$               & $200\pm\mathbf{100}$  & $1.0\times10^{-4}$       & $6.7\times10^{-5}$ \\
  wide range    & $3\times10^{-6}$                & $30$               & $-1.0$               & $200\pm\mathbf{1000}$   & $1.3\times10^{-4}$       & $13.8\times10^{-5}$ \\
  short delay   & $3\times10^{-6}$                & $\mathbf{10}$ & $-1.0$                & $200\pm500$                  & $1.3\times10^{-4}$      & $6.8\times10^{-5}$ \\
  steep DTD    & $3\times10^{-6}$                & $30$               & $\mathbf{-1.5}$ & $200\pm500$                   & $0.054\times10^{-4}$     & $27.5\times10^{-5}$ \\
\hline \\[-4mm]                                                                                                                                       
  0.1\% of all ccSNe   & & & & & $0.37\times10^{-4}$ & $7.3\times10^{-5}$ \\
  0.01\% of all ccSNe & & & & & $0.039\times10^{-4}$ & $75.1\times10^{-5}$ \\
\hline                                                                                                                                                
\end{tabular}                                                                                                                                         
\end{center}                                                                                                                                          
\end{table*}      

In this work, we study the impact of introducing neutron star natal kicks on the resulting r-process abundance ratios in a Milky Way-mass system. In previous cosmological simulations of Milky Way-mass galaxies, r-process elements were released around the star particle (representing a single stellar population) in which the neutron star binary originally formed, thus neglecting natal kicks altogether \citep[e.g.][]{Shen2015, Voort2015, Naiman2018, Haynes2019, Voort2020}. Here, we include natal kicks by generating additional massless neutron star merger particles which are given an additional velocity kick in a random direction. The original star particles are treated in the same way as before. When a star particle forms, we stochastically determine whether or not its stellar population will produce one or more neutron star mergers in the future, based on the following rate:
\begin{equation} \label{eqn:rate}
  R_\mathrm{NS}=AM_\star t^\gamma\ \mathrm{for}\ t>t_\mathrm{min}
\end{equation}
and $R_\mathrm{NS}=0$ when $t<t_\mathrm{min}$. Here, $A$ is the number of neutron star mergers per unit of stellar mass, $M_\star$ is the mass of the star particle formed, $t$ is the time since the formation of the star particle, $\gamma$ is the (always negative) exponent of the time dependence of the delay time distribution, and $t_\mathrm{min}$ is the minimum time needed for a neutron star merger to occur \citep[e.g.][]{Belczynski2006}. We use the following values for our fiducial neutron star merger model: $A=3\times10^{-6}$~M$_\odot^{-1}$, $\gamma=-1$, and $t_\mathrm{min}=3\times10^7$~yr. These are the same values we used before in \citet{Voort2020} and were chosen as a reasonable guess based on the available literature \citep{Belczynski2006, Zheng2007, Abadie2010, Berger2014, Avanzo2015, Cote2019, Kim2015, Abbott2017}. In Section~\ref{sec:optimal} we will use parameter values that were chosen to produce a better match with available observations of stellar abundances, but still lie within current observational and theoretical uncertainties ($A=3\times10^{-5}$~M$_\odot^{-1}$, $\gamma=-1.5$, and $t_\mathrm{min}=10^7$~yr).

If a star particle (i.e.\ a stellar population) will produce one or more neutron star mergers before the end of the simulation at $z=0$, an equivalent number of neutron star merger particles are generated. These particles are massless, which is a good approximation, since a binary neutron star system is much less massive than our star particles. The neutron star merger particles have the same velocity as the star particle from which they spawned plus an additional kick in a random direction. The kick velocity is chosen from a Gaussian distribution, with a peak at 200~km~s$^{-1}$ and a standard deviation of 500~km~s$^{-1}$ for our fiducial kick model.\footnote{Because the direction is chosen separately and sampled isotropically, we discard negative velocities and draw a new value from the distribution.} We include multiple kick models in the same simulation with different kick velocities, because the values are highly uncertain, especially for binary systems. We can use a single cosmological simulation to study multiple r-process models, because they do not impact the rest of the simulation and behave as passive tracers (see Section~\ref{sec:sim}). The resulting r-process elements produced are stored in separate variables for each model. The parameter values used are listed in Table~\ref{tab:models}. Note that model `no kick' produces neutron star merger particles with an initial velocity equal to that of their parent stellar population, but without an additional kick, and is therefore identical to the fiducial model in \citet{Voort2020} even though the implementation is slightly different.

Table~\ref{tab:models} also lists the resulting $z=0$ rate of r-process producing events (either by neutron star mergers or by a rare type of core-collapse supernova) in the entire zoom-in region of the simulation. This corresponds to a maximum distance of approximately 1~Mpc from the central galaxy. However, the r-process events are strongly centrally concentrated, so reducing the maximum distance only changes the rates slightly. Although the first 6 models have the same delay time distribution, $z=0$ rates vary somewhat due to the stochasticity involved in randomly sampling this distribution, which is done independently for each model. The last column gives the resulting europium yield obtained by setting the median $\mathrm{[Eu/Mg]}$ to zero at $\mathrm{[Mg/H]}=0$. The rates and yields of our models are consistent with most observational constraints within uncertainties \citep[e.g.][]{Abadie2010, Behroozi2014, Kim2015, Abbott2017}, though our rates are lower than the estimate from \citet{Pol2019} who found $R_\mathrm{NS}(z=0)=42^{+30}_{-14}$~Myr$^{-1}$. Note that the model with a steep exponent for the delay time distribution has a low-redshift event rate more than an order of magnitude lower than the other models, because fewer neutron star mergers occur as the stellar population ages. We will revisit this in Section~\ref{sec:optimal}, which discusses results from a separate medium-resolution simulation with parameters chosen to better match observations of stellar r-process abundances.

We have performed a resolution test by running identical simulations with 8 and 64 times lower mass resolution. The models tested, both with and without natal kicks, are almost perfectly converged. The convergence behaviour is very similar to that shown in \citet{Voort2020}. As in our previous work, there is a very small residual resolution dependence. The median r-process abundance ratio decreases marginally and the scatter increases slightly as the resolution improves. This small difference can likely be explained by the decrease of numerical mixing with increasing resolution. Because these differences are so minor, we do not believe they are the main uncertainty associated with this work. Higher resolution does substantially increase the statistics by increasing the number of star particles and therefore allows us to probe down to lower metallicities.

\subsection{Modeling rare core-collapse supernovae} \label{sec:SN}

Although the main topic of this work is the effect of adding neutron star natal kicks, we also explore the r-process enrichment from a rare type of core-collapse supernova for comparison. Candidates for such sources include collapsars and magneto-rotational supernovae \citep[see e.g.][]{Cote2019}, but we remain agnostic towards the exact source and keep our models as simple as possible. As was done for the neutron star mergers, r-process events are sampled stochastically and each r-process producing core-collapse supernova releases the same amount of r-process elements. The resulting stellar abundance ratios are normalized by choosing the europium yield such that the median $\mathrm{[Eu/Mg]}$ is zero at $\mathrm{[Mg/H]}=0$ in post-processing, as given in the last column of Table~\ref{tab:models}.

In our two models included here, either 1 in 1,000 (0.1 per cent) or 1 in 10,000 (0.01 per cent) core-collapse supernovae act as sources of r-process elements \citep{Beniamini2016}. These are not modeled with massless particles as was done for the neutron star merger models. Their enrichment rather occurs directly in the host cell of the original star particle containing the core-collapse supernova. This is slightly different to their treatment in \citet{Voort2020} where r-process elements were released into $\approx64$ neighbouring cells, but this is unlikely to affect the results, as also shown and discussed in \citet{Voort2020}. From Table~\ref{tab:models} we can see that the 1 in 1,000 model has an event rate about a third of our fiducial neutron star merger model. The rate is another full order of magnitude lower for the model with only 1 in 10,000 core-collapse supernovae as r-process production sites.

\section{Results}\label{sec:results}

In this Section, we show results for all stars within the virial radius of our simulated Milky Way-mass galaxy at $z=0$, including disc stars as well as halo stars. To calculate the median and scatter in r-process abundance ratios, we use 0.2~dex metallicity bins and show only those bins that contain at least 100 star particles. The exception is Figure~\ref{fig:rpvel} where we reduced the minimum number of star particles to 50 in order to probe to lower metallicities at the expense of increased noise in those bins. In this work, for brevity, we refer to the 16th and 84th percentiles as the 1$\sigma$ scatter and to the 2nd and 98th percentile of the distribution as the $2\sigma$ scatter, even when the distribution is not Gaussian.

\subsection{Neutron star mergers with natal kicks}\label{sec:kicks}

\begin{figure}
\center
\includegraphics[scale=0.45]{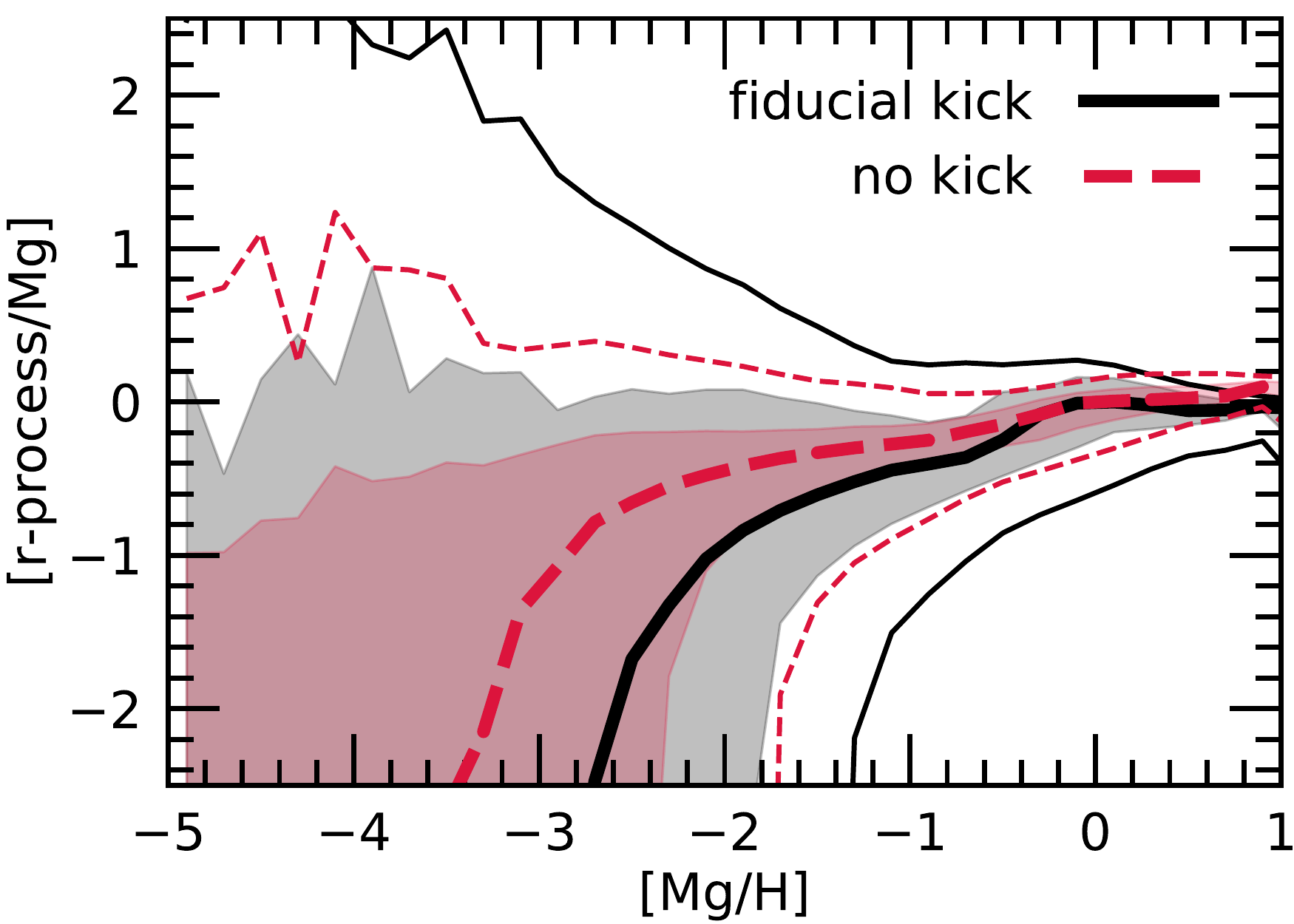}
\caption {\label{fig:rpkick} The abundance ratio $\mathrm{[\rp/Mg]}$ as a function of metallicity ($\mathrm{[Mg/H]}$) for our fiducial neutron star merger model with $v_\mathrm{kick}=200\pm500$~km~s$^{-1}$ in black (solid curves) and the same model without natal kicks in red (dashed curves). Thick curves show the median relation, while the thin curves show the $2\sigma$ scatter. The grey and red shaded regions cover the $1\sigma$ scatter. When kicks are included, the median is lower and the increase of $\mathrm{[\rp/Mg]}$ with metallicity is steeper. The $1\sigma$ and $2\sigma$ scatter in both directions are substantially larger with natal kicks.} 
\vspace{-3mm}
\end{figure} 

We first focus on the difference in stellar abundance ratios between our fiducial model with natal kicks included (`fiducial kick') and a model with identical paramaters except for setting all kick velocities to zero (`no kick'). In Figure~\ref{fig:rpkick}, we show $\mathrm{[\rp/Mg]}$ as a function of metallicity for all stars within $R_\mathrm{vir}$ at $z=0$. The thick curves show the median, the shaded region show the 16th and 84th percentiles (from now on refered to as the 1$\sigma$ scatter), and the thin curves show the 2nd and 98th percentiles (from now on refered to as the 2$\sigma$ scatter). Our fiducial kick model is shown as solid, black curves and the model without kicks as dashed, red curves.

There are two very clear differences between the two enrichment models. First, the inclusion of natal kicks lowers the median r-process abundances at $\mathrm{[Mg/H]}<-0.5$ and the difference between the two models becomes larger towards lower metallicities. Because we normalize our models to $\mathrm{[\rp/Mg]}=0$ at $\mathrm{[Mg/H]}=0$, the r-process yields used are slightly different depending on whether or not natal kicks are included. As can be seen from Table~\ref{tab:models}, in the `fiducial kick model' we use yields that are 50 per cent larger than in the model without kicks. If we had used the same normalization, the solid, black `fiducial kick' curve would shift down by 0.17 dex and lie below the dashed, red `no kick' curve at all metallicities. This reduction of the stellar r-process abundances when kicks are included was expected, because neutron star mergers happen, on average, further away from the galaxy. This does not necessarily mean that the produced elements are lost to the galaxy, because they can accrete onto the ISM at a later time and become incorporated in subsequent stellar populations. However, the fraction of r-process elements escaping the system is likely larger when they are produced at larger distances. Note that the normalization difference is fairly minor, which indicates that the accretion of r-process elements is important. 

\begin{figure*}
\center
\includegraphics[scale=0.45]{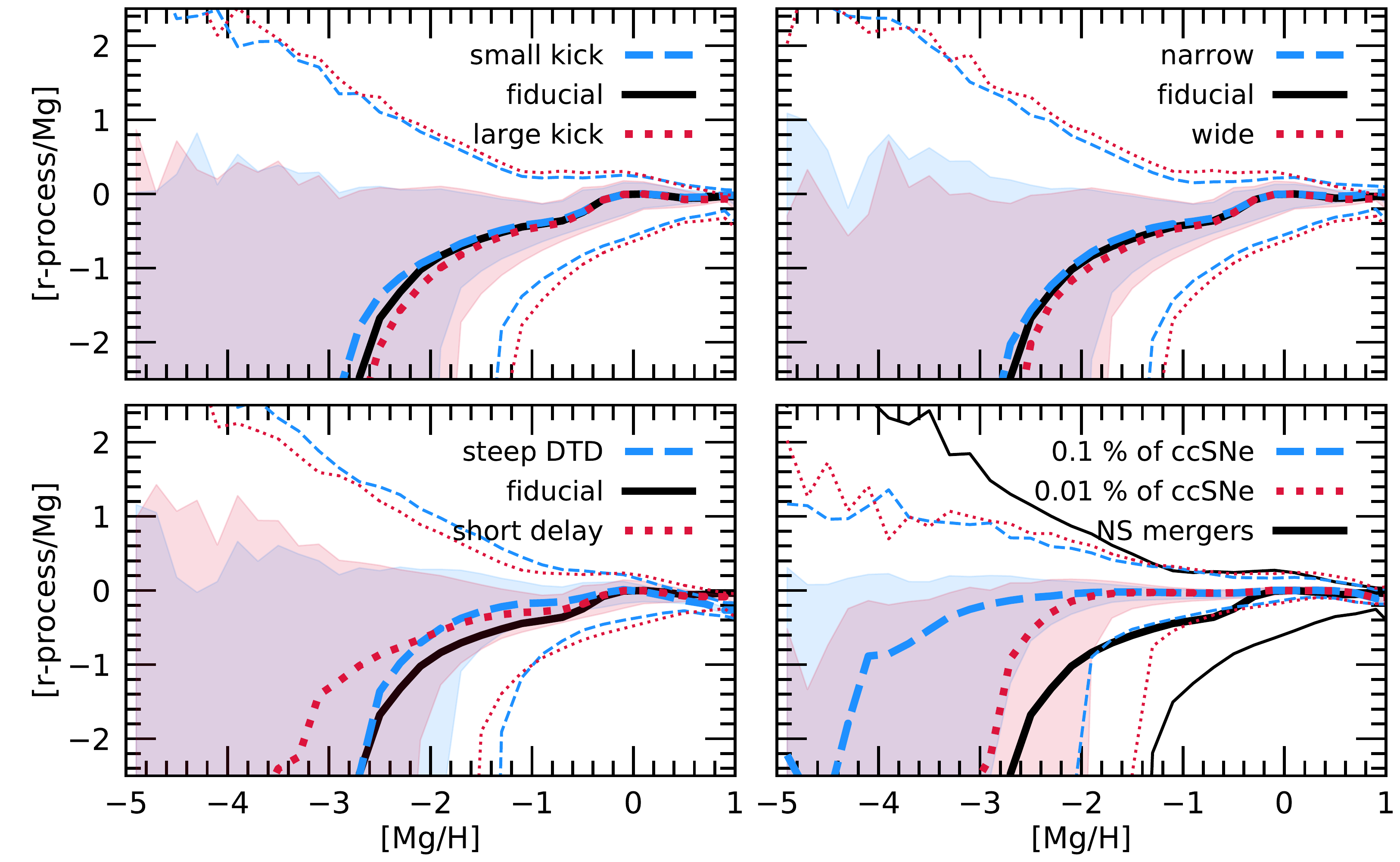}
\caption {\label{fig:rpvar} $\mathrm{[\rp/Mg]}$ as a function of $\mathrm{[Mg/H]}$ for a variety of models with different parameters as listed in Table~\ref{tab:models}. Thick curves show the median value, whilst the shaded regions cover the 1$\sigma$ scatter and the thin curves show the $2\sigma$ scatter. The two top panels compare variations on the kick velocities of the models and show almost identical results. This means that the model is insensitive to the exact value for the kick velocity distribution chosen, though the average velocity in all our models is at least a few hundred~km~s$^{-1}$. The bottom left-hand panel explores variations on parameters of the neutron star merger delay time distribution. A steeper time dependence and a shorter delay increase the r-process abundance at $\mathrm{[Mg/H]}<1$. The bottom right-hand panel includes the two rare core-collapse supernova models. For these, the trend of $\mathrm{[\rp/Mg]}$ with metallicity is flatter and the scatter is smaller than for our neutron star merger models. All our models show a median r-process abundance that drops at low-metallicity due to the source being too rare to enrich the ISM in the early Universe.} 
\vspace{-3mm}
\end{figure*} 

The second substantial difference is that the scatter in r-process abundance is much larger when natal kicks are included, towards both high and low values of $\mathrm{[\rp/Mg]}$. This is because a neutron star binary can travel far from its birth site before merging.\footnote{A binary moving at~200 km s$^{-1}$ will travel 6~kpc in 30 Myr and 200~kpc in a Gyr.} They are thus more likely to release r-process elements in an area that is likely not to have seen star formation itself and therefore has not been enriched directly by core-collapse supernovae, although it may have been polluted by large-scale galactic outflows. Only 1 (0.5) per cent of stars have $\mathrm{[\rp/Mg]}>1$ at $-5<\mathrm{[Mg/H]}<-3$ ($-5<\mathrm{[Mg/H]}<-2$) without kicks, but this fraction rises by an order of magnitude to 10 (4) per cent when kicks are included. Extreme outliers in observations are therefore more easily reproduced in a model where the binaries receive a substantial velocity kick. 

At the highest metallicities, another difference can be seen, although this is a very minor effect. The model with natal kicks decreases slightly towards high metallicity whilst the model without kicks continues to increase slightly. The difference between the models is only 0.1~dex and may therefore not be important, but we discuss it here briefly. The likely reason for this difference in behaviour is linked to the star formation history (SFH) of this galaxy. Its SFH increases from its formation up to a look-back time of 9~Gyr after which it fluctuates around 15~M$_{\astrosun}$~yr$^{-1}$ for 5~Gyr and decreases over the last 4~Gyr to approximately 3~M$_{\astrosun}$~yr$^{-1}$. This means that prompt neutron star mergers become less frequent towards the present day and therefore neutron star mergers with long delay times become relatively more important. It is these long delay time systems that spend a substantial time traveling away from the galaxy, when natal kicks are included, and merge far away from its ISM, thus enriching the intergalactic medium. This results in a lowering of the r-process to magnesium abundance ratio at late times (i.e.\ at high metallicity) as magnesium is produced by supernovae inside (or close to) the ISM.

\subsection{Parameter study}\label{sec:var}


The parameter values used in Section~\ref{sec:kicks} are highly uncertain. We therefore included several additional r-process enrichment models based on neutron star mergers, where for each model we varied a single parameter of either the kick velocity distribution or the delay time distribution. These values are summarized in Table~\ref{tab:models}. We note that model `wide range' contains stars with kick velocities higher than observed \citep{Hobbs2005, Faucher2006}, but it is nevertheless included to understand the effect of increasing the width of the kick distribution. We also added two models where rare core-collapse supernovae produce all of the r-process elements (as described in Section~\ref{sec:SN}). The models included here are not exhaustive, because we only vary a single parameter each time. Further models are explored via a medium-resolution simulation in Section~\ref{sec:optimal}. Figure~\ref{fig:rpvar} shows the resulting r-process abundance ratios as a function of metallicity. In each panel, the thick, black curve shows the median relation for our fiducial neutron star merger model (identical to the `fiducial kick' model in Figure~\ref{fig:rpkick}). Dashed, blue and dotted, red curves show the median ($2\sigma$ scatter) as thick (thin) curves and the $1\sigma$ scatter as shaded regions for the model variations.

Overall, the differences between our neutron star models with natal kicks are relatively mild (top panels and bottom, left-hand panel) and much smaller than the difference between models including or excluding natal kicks (see Figure~\ref{fig:rpkick}). The small differences that are present behave as expected. Decreasing (increasing) the average kick velocity marginally increases (decreases) the median $\mathrm{[\rp/Mg]}$ at low metallicity, flattening the relation with metallicity slightly, and decreases (increases) the scatter (top, left-hand panel). Decreasing (increasing) the width of the kick velocity distribution shows mostly the same behaviour (top, right-hand panel). However, the $1\sigma$ scatter is higher at low-metallicity for the narrower kick distribution model. This is likely because more neutron star mergers occur inside or close to the progenitor galaxies in the early Universe, enriching their ISM more easily. Decreasing the minimum delay time and steepening the delay time distribution increase the median $\mathrm{[\rp/Mg]}$ because neutron star mergers occur earlier, on average (bottom, left-hand panel; as also found in \citealt{Voort2020}). These changes have a somewhat larger effect than the kick velocity variations we tested here.

For all of our models shown in Figure~\ref{fig:rpvar}, we renormalized their abundances to $\mathrm{[\rp/Mg]}=0$ at $\mathrm{[Mg/H]}=0$, effectively changing the r-process yield in post-processing. The last column in Table~\ref{tab:models} lists the yields used and from this we can see how using a fixed yield would increase the difference between models. For the models with the centroid of the kick velocity distribution shifted, the normalization only changes by 0.14~dex between model `small kick' and model `large kick'. However, changing the width of the distribution from 100 to 1000~km~s$^{-1}$ decreases $\mathrm{[\rp/Mg]}$ by 0.31~dex if we were to use the same yields. This is because neutron star binaries traveling at very high velocities are likely to merge so far away from the galaxy that the elements they produce will not be able to accrete onto the galaxy and are therefore not incorporated into future generations of stars. The normalization of model `steep DTD' is the most discrepant, 0.45~dex higher than our fiducial model. With a steeper time dependence, fewer neutron star mergers occur for a given stellar population and the difference increases with time. If we were to use fixed yields, model `steep DTD' would result in lower r-process abundances at high metallicity compared to the fiducial model. 

To allow direct comparison to models with alternative sources for r-process enrichment, we included two models where a special class of core-collapse supernovae, such as collapsars or magneto-rotational supernovae, is responsible for the synthesis of all r-process elements. These are shown in the bottom, right-hand panel of Figure~\ref{fig:rpvar}. We use an event rate of either 1 in every 1,000 core-collapse supernovae (dashed, blue curves and blue shaded area) or 1 in every 10,000 (dotted, red curves and red shaded area). The median relation between $\mathrm{[\rp/Mg]}$ and $\mathrm{[Mg/H]}$ is flat at high and intermediate metallicity. Towards very low metallicity, the median r-process abundance ratio decreases. As expected, this decrease occurs at a higher metallicity for a source that is more rare (i.e. 1 in 10,000 core-collapse supernovae) because it is difficult to enrich the entire ISM in the early universe when the event rate is that low. This is similar to the steep decline of $\mathrm{[\rp/Mg]}$ towards very low metallicity in neutron star merger models. 

For both rare core-collapse supernova models, the upper $1\sigma$ bound stays close to $\mathrm{[\rp/Mg]}=0$ at any metallicity. At $\mathrm{[Mg/H]}\gtrsim-2$, the $1\sigma$ scatter is very small and rapidly increases to lower metallicites. The scatter is smaller than that in any of the neutron star merger models with natal kicks shown in the other panels. To quantify this somewhat, only 2 (1) per cent of stars have $\mathrm{[\rp/Mg]}>1$ at $-5<\mathrm{[Mg/H]}<-3$ ($-5<\mathrm{[Mg/H]}<-2$) in the two rare core-collapse supernova models, compared to the various neutron star merger models with natal kicks presented here for which the fractions of similarly r-process enhanced stars vary between 9 and 14 (4 and 6) per cent. Future observations of a homogeneous sample of (extremely) metal-poor stars would be able to measure the fraction of stars with high r-process abundance ratios, which could then be compared to the values obtained from our different models.

\subsection{Optimized neutron star merger model with natal kicks}\label{sec:optimal}

\begin{table*}
\begin{center}                                                                                                                                        
\caption{\label{tab:optimal} \small Parameters of r-process enrichment models for those based on neutron star mergers which was optimized to match available observations. The columns list the same properties as Table~\ref{tab:models}.}  
\begin{tabular}[t]{llrrrrr}
\hline \\[-3mm]                                                                                                                                       
model name   & $A$                                   & $t_\mathrm{min}$              & $\gamma$           & $v_\mathrm{kick}$  & $R_\mathrm{\rp}(z=0)$ & $y_\mathrm{Eu}$ \\
                      & (M$_{\astrosun}^{-1}$)                  & (Myr)                                  &                           & (km s$^{-1}$)      & (yr$^{-1}$)             & (M$_{\astrosun}$)\\
\hline \\[-4mm]                                                                                                                                       
  optimized; fiducial kick                            & $3\times10^{-5}$                & $10$               & $-1.5$               & $200\pm500$                           & $7.1\times10^{-5}$       & $2.7\times10^{-5}$ \\
  optimized; small kick                               & $3\times10^{-5}$                & $10$               & $-1.5$               & $\mathbf{0}\pm500$                 & $7.0\times10^{-5}$       & $2.7\times10^{-5}$ \\
  optimized; small kick; narrow range        & $3\times10^{-5}$                & $10$               & $-1.5$               & $\mathbf{0}\pm\mathbf{100}$  & $6.8\times10^{-5}$       & $2.8\times10^{-5}$ \\
  optimized; small kick; narrower range     & $3\times10^{-5}$                & $10$               & $-1.5$               & $\mathbf{0}\pm\mathbf{50}$    & $6.8\times10^{-5}$       & $2.8\times10^{-5}$ \\
  optimized; no kick                                   & $3\times10^{-5}$                & $10$               & $-1.5$               & $\mathbf{0}\pm\mathbf{0}$      & $6.7\times10^{-5}$       & $2.9\times10^{-5}$ \\
\hline                                                                                                                                                
\end{tabular}                                                                                                                                         
\end{center}                                                                                                                                          
\end{table*}      

The increasing trend of $\mathrm{[\rp/Mg]}$ (or similarly $\mathrm{[\rp/Fe]}$) with metallicity for all our neutron star merger models in Sections~\ref{sec:kicks} and~\ref{sec:var} is not seen in observations (see \citealt{Voort2020} for a more detailed comparison and discussion of our fiducial model without kicks). We note that the observational data for extremely metal-poor stars is sparse and could be biased. Nevertheless, currently available observations seem to prefer a flat trend of $\mathrm{[\rp/Mg]}$ with metallicity, which is easier to achieve with a model based on rare (but prompt) core-collapse supernovae as r-process production sites (see also Section~\ref{sec:var} and Figure~\ref{fig:rpvar}).

Until now, we attempted no tuning of our neutron star merger parameters. Motivated by the inability to reproduce observational trends, we resimulated the same galaxy at medium resolution (i.e.\ 8 times lower mass resolution than our fiducial simulation) with new neutron star delay time distribution values specifically chosen to improve the match with observations: a higher delay time distribution normalization, a shorter minimum delay time, and a steeper exponent (see Table~\ref{tab:optimal}). With these new delay time parameters, we varied the kick velocities and refer to this new set of neutron star models as `optimized'.  The exponent of the delay time distribution is set to $\gamma=-1.5$ as in model `steep DTD', the minimum delay time is reduced to 10~Myr as in model `short delay', and the normalization is increased by an order of magnitude. Because of the steepness of the distribution, the change in normalization is not as extreme as it may seem: the low-redshift neutron star merger rate is below that of our fiducial model with $\gamma=-1$ and in agreement with or even below rate estimates from the literature \citep[e.g.][]{Abadie2010, Pol2019}. 

\begin{figure}
\center
\includegraphics[scale=0.45]{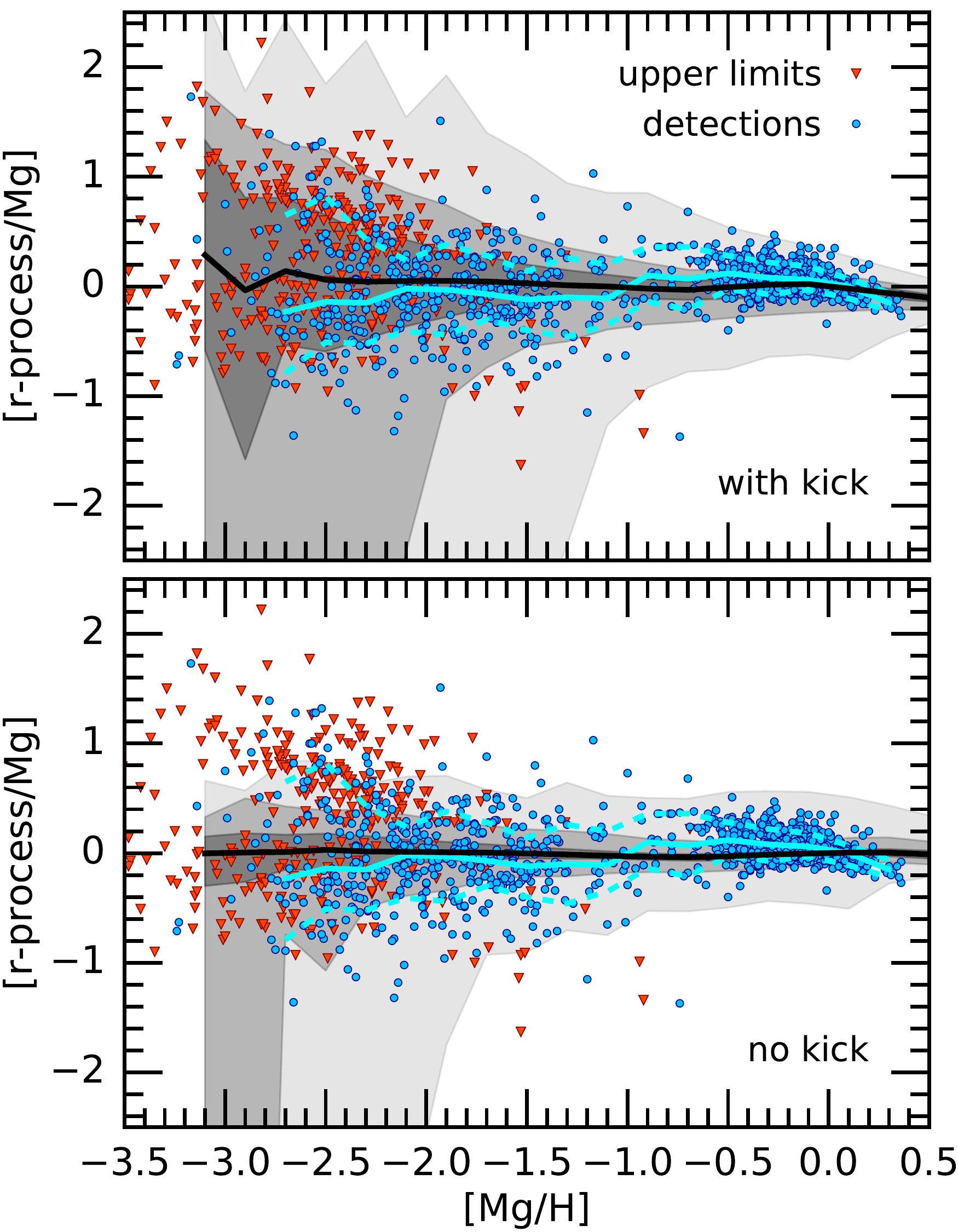}
\caption {\label{fig:rpobs} Median (solid, black curves) and $1\sigma$, $2\sigma$, and $3\sigma$ scatter (grey shaded regions) of $\mathrm{[\rp/Mg]}$ as a function of $\mathrm{[Mg/H]}$ for two neutron star merger models with optimized parameters. One model includes our fiducial kick velocity ($v_\mathrm{kick}=200\pm500$~km~s$^{-1}$; top panel) and the other one does not include natal kicks (bottom panel). Observations of $[\mathrm{Eu/Mg}]$ in Milky Way stars are shown as blue circles (detections) and red downward triangles (upper limits) extracted from the SAGA database compiled by \citet{Suda2008}. The solid cyan lines show the median and the dashed, cyan curves show the $1\sigma$ scatter of the observational detections. The two models reproduce the mostly flat behaviour of the r-process abundance ratio equally well. However, the model without kicks vastly underproduces the scatter seen in observations, whereas the match between observations and the model with kicks is excellent at low metallicity. In this `optimized' model, the scatter is determined by the neutron star kick velocities rather than by the rarity of r-process sources.} 
\vspace{-3mm}
\end{figure} 

Figure~\ref{fig:rpobs} shows the abundance ratios from two of our `optimized' neutron star merger models. The top panel shows the median (solid, black curve) and $1\sigma$, $2\sigma$, and $3\sigma$ scatter (grey shaded regions) of $\mathrm{[\rp/Mg]}$ in our fiducial kick velocity distribution with $v_\mathrm{kick}=200\pm500$~km~s$^{-1}$ and the bottom panel shows the same for a model without natal kicks. The `optimized' parameters, chosen within the range allowed by current constraints, result in higher average r-process abundances in metal-poor stars compared to our fiducial parameters (see Section~\ref{sec:kicks}). The relation between the median $\mathrm{[\rp/Mg]}$ and $\mathrm{[Mg/H]}$ is mostly flat, whereas there was a rising trend with metallicity in our fiducial model shown in Figure~\ref{fig:rpkick}.

It is important to point out that the median is very similar for the two models shown and therefore is not strongly affected by natal kicks in this case. This is different from the behaviour shown in Section~\ref{sec:kicks} for our fiducial delay time distribution parameters, where the addition of kicks reduced the median $\mathrm{[\rp/Mg]}$. Comparing the adopted europium yields in Table~\ref{tab:optimal} for our fiducial kick model and the one without kicks confirms there is little difference in the normalization. Neutron star mergers occur more rapidly after formation of the stellar population in the `optimized' model than in our fiducial model. Therefore, the binary travels less far from its birth position, which decreases the impact of natal kicks. 

Looking at the scatter around the median, we see that it is much smaller than in our models with fiducial delay time distribution parameters, both with and without kicks, as shown in Section~\ref{sec:kicks}. Qualitatively, this is expected, because the number of neutron star mergers is higher and the average delay time for neutron star mergers is shorter. The rate of mergers at $z=0$ is lower than for our original model -- compare Tables~\ref{tab:models} and~\ref{tab:optimal} -- but the rate is higher at high redshift. In the early universe, at $z>3$, the number of neutron star mergers is 4 times higher in our `optimized' model. Over the full 13.8~Gyr of simulation time, the `optimized' model produces 60 per cent more neutron star mergers than the fiducial model. Furthermore, the fact that the neutron mergers occur more promptly means that they occur closer in time and space to core-collapse supernovae, which produce magnesium. These two differences explain the decreased scatter in $\mathrm{[\rp/Mg]}$.

When comparing the scatter in our `optimized' model with kicks to our `optimized' model without kicks, the same conclusion holds as before: neutron star natal kicks substantially increase the scatter in r-process abundance, especially at low metallicity. The scatter is clearly dominated by the presence of neutron star kicks. This has important implication for the comparison of our models to observations.

\subsubsection{Comparison to observations}\label{sec:obs}

We now wish to see whether our `optimized' model with neutron star mergers as the only source of r-process elements can match currently available observations of stellar abundances. Besides abundances from our `optimized' models, Figure~\ref{fig:rpobs} also shows observational detections (blue circles) and upper limits (red downward triangles) from the Stellar Abundances for Galactic Archeology (SAGA) database \citep{Suda2008}. The median (solid, cyan curve) and $1\sigma$ scatter (dashed, cyan curves) of the observational data (detections only) are included if the 0.2~dex metallicity bins contained at least 10 data points. Note that the observational sample is not homogeneous and could suffer from selection bias. Also note that our simulations do not have any measurement errors included, unlike observations. No exact match is expected, but it is still illuminating to compare our simulation results to the available observational data.

The median $\mathrm{[\rp/Mg]}\approx0$ at all metallicities in the `optimized' model, both with and without natal kicks. This is similar to the behaviour in observations and what we were aiming for when choosing new parameters for the delay time distribution. The main difference between including or excluding natal kicks lies in the scatter around the median. It is clear that the model with kicks much better reproduces the $1\sigma$ observational scatter as well as extreme outliers than the model without kicks.

The conclusion that this model with `optimized' parameters and $v_\mathrm{kick}=200\pm500$~km~s$^{-1}$ provides a reasonable match to currently available observations does not necessarily imply it is the correct one. First, a larger and unbiased observational survey is needed to improve statistics for metal-poor stars and sample the population in a homogeneous way. Second, there may be other possible neutron star merger parameter choices that would also result in a r-process element distribution similar to observations. Similarly, there may be other viable sources, such as a special type of core-collapse supernova, able to match observations as well. However, what our results prove is that it is possible to create an r-process enrichment model with reasonable parameters using neutron star mergers as the only source of r-process elements that matches available data.

It is important to point out that in our `optimized' model, the scatter in r-process abundances is created by the neutron star kicks, which decouple the location of core-collapse supernovae and neutron star mergers more to a much larger extent than a model without kicks. This is a clear departure from the usual interpretation, which is that the large scatter at low metallicity is caused by inhomogeneous enrichment due to the rarity of the r-process source. This possibility should be kept in mind when interpreting observational data.

\subsubsection{Lower velocity natal kicks}\label{sec:vel}

In Section~\ref{sec:var} we showed that for moderate changes to the kick velocity, the resulting abundance ratios do not vary substantially. However, all our models had relatively high average velocities of at least a few hundred km~s$^{-1}$. Although this is still debated, there is evidence of lower kick velocities in neutron star binaries \citep[e.g.][]{Fong2013, BeniaminiPiran2016, Giacobbo2018}. If the velocity is lowered enough, the abundances should resemble the model without kicks. Therefore we included a few additional models in our medium-resolution simulation with significantly reduced kick velocities. The parameter values of the kick distributions in these new model variations are detailed in Table~\ref{tab:optimal}.

\begin{figure}
\center
\includegraphics[scale=0.45]{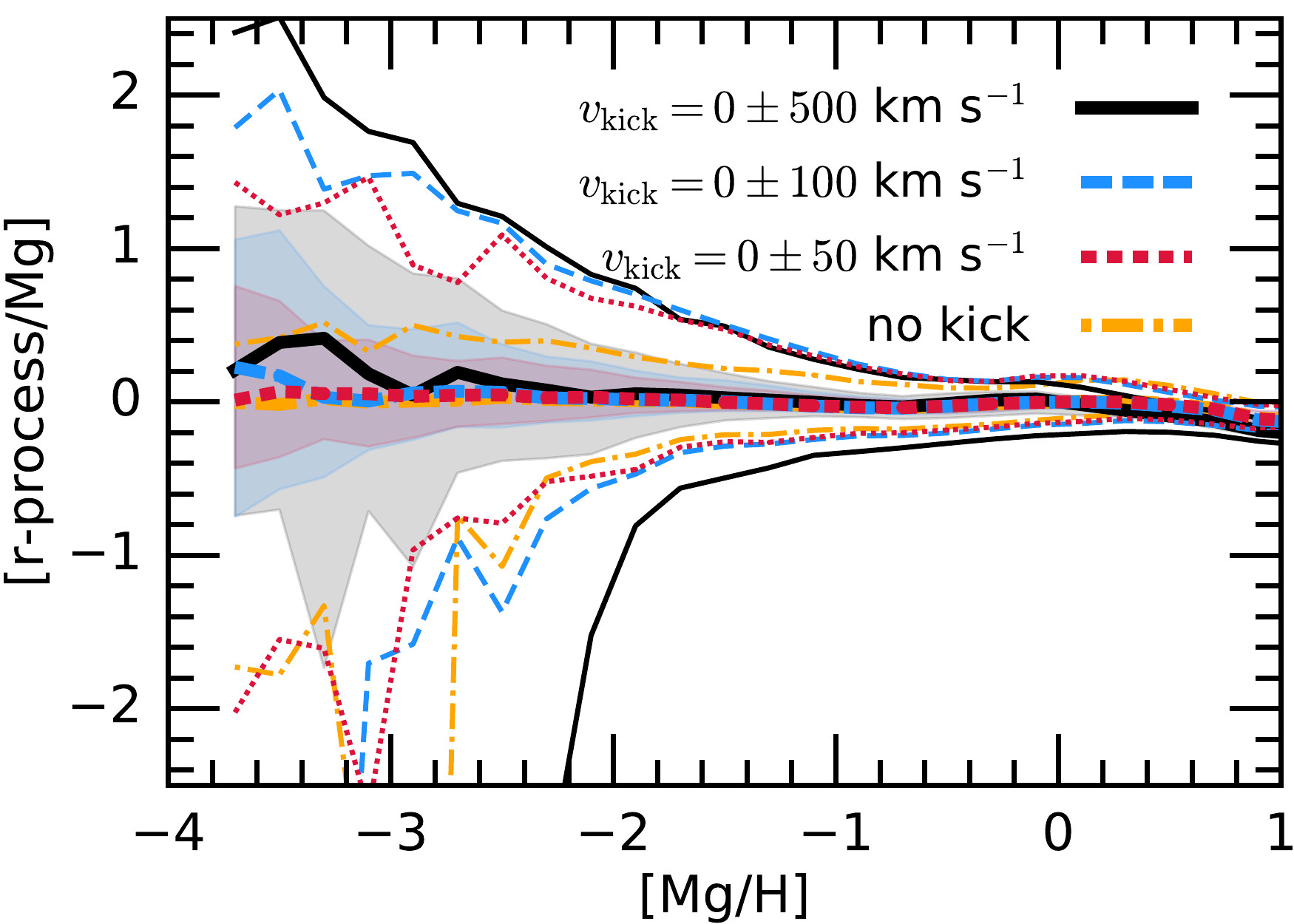}
\caption {\label{fig:rpvel} $\mathrm{[\rp/Mg]}$ as a function of $\mathrm{[Mg/H]}$ for neutron star merger models with different kick velocities as listed in Table~\ref{tab:optimal}. Thick curves show the median value, while the shaded regions and thin curves indicate the $1\sigma$ and $2\sigma$ scatter, respectively. The model without kicks is shown as dot-dashed, orange curves and is identical to the model in the bottom panel of Figure~\ref{fig:rpobs}. The other models all include natal kicks picked from a Gaussian distribution centred at zero and with a standard deviation decreasing from 500 (solid, black curves) to 100 (dashed, blue curves) to 50~km~s$^{-1}$ (red, dotted curves). The scatter decreases as the average kick velocity decreases, but not by the same amount for the upwards and downwards scatter. The lower $2\sigma$ bound for the model with $v_\mathrm{kick}=0\pm100$~km~s$^{-1}$ follows that of the model without kicks, whereas its upper $2\sigma$ bound is very close to the model with much larger kicks of $v_\mathrm{kick}=0\pm500$~km~s$^{-1}$. Even the model with a modest $v_\mathrm{kick}=0\pm50$~km~s$^{-1}$ has a much larger scatter than the model without kicks. Thus, even relatively low velocity kicks have a substantial impact on the resulting scatter in r-process abundances.} 
\vspace{-3mm}
\end{figure} 

The resulting r-process abundance ratios are shown in Figure~\ref{fig:rpvel} for 4 neutron star merger models with kick velocity distributions centred at 0 and a standard deviation of 500~km~s$^{-1}$ (solid, black curves and grey shaded region), 100~km~s$^{-1}$ (dashed, blue curves and shading), 50~km~s$^{-1}$ (dotted, red curves and shading), and 0~km~s$^{-1}$ (dot-dashed, orange curves). The latter model does not include kicks. This model is identical to the simulation shown in the bottom panel of Figure~\ref{fig:rpobs} and its shading has been omitted for clarity. Thick and thin curves show the median and $2\sigma$ scatter, respectively, and the shaded areas show the $1\sigma$ scatter. To probe to slightly lower metallicities, we included metallicity bins that include at least 50~star particles instead of our fiducial 100 used in all other figures. The statistics are therefore a bit worse than in our other figures for $\mathrm{[Mg/H]}<-3.1$.

As we also saw in Figure~\ref{fig:rpobs}, reducing the kick velocity in our `optimized' model results in a fairly similar median relation of $\mathrm{[\rp/Mg]}$ versus $\mathrm{[Mg/H]}$. There is a hint of a small decrease in $\mathrm{[\rp/Mg]}$ at low metallicity for lower velocities as compared to the $v_\mathrm{kick}=0\pm500$~km~s$^{-1}$ model, but this would need to be confirmed by high-resolution simulations with better statistics. At high metallicities, the lower velocity models have a slightly higher $\mathrm{[\rp/Mg]}$ and therefore a slightly flatter metallicity dependence, but this difference is also very minor.

In contrast to our previous moderate changes in the kick velocity distribution (see Figure~\ref{fig:rpvar} and Section~\ref{sec:var}), the scatter in r-process abundances is clearly affected when reducing the average kick velocity to less than a hundred~km~s$^{-1}$.\footnote{The mean (median) absolute value of a Gaussian distribution centred on zero with a standard deviation of 100~km~s$^{-1}$ is 80~(67)~km~s$^{-1}$.} The $v_\mathrm{kick}=0\pm100$ and $0\pm50$~km~s$^{-1}$ each have lower scatter than the $v_\mathrm{kick}=0\pm500$~km~s$^{-1}$ model. However, their scatter is still larger than in the model without natal kicks (see Figure~\ref{fig:rpobs} for the $1\sigma$ scatter of model `no kick'). The $2\sigma$ scatter above the median relation in the low-velocity models is almost as high as in the high-velocity kick models, so r-process enhanced outliers will still be relatively common, whereas their $2\sigma$ scatter below the median is more similar to the model without kicks. The overall conclusion is that the effect of natal kicks is reduced when their average $v_\mathrm{kick}\lesssim100$~km~s$^{-1}$ as compared to our fiducial model. However, even these relatively low-velocity kicks still increase the scatter in stellar r-process abundances as compared to a model without kicks.

\section{Discussion and conclusions} \label{sec:concl}

We explored the stellar abundances of r-process elements in cosmological, magnetohydrodynamical simulations of a Milky Way-mass galaxy using the moving mesh code \textsc{arepo} using the Auriga galaxy formation model. We implemented a variety of models for r-process enrichment, most of which use neutron star mergers as the only source of r-process elements, though two use rare core-collapse supernovae as sole r-process production sites. In this work, we focused on understanding the impact of adding neutron star natal kicks to the neutron star merger models. Natal kicks cause neutron star binaries to move far from their birth positions, which changes the resulting chemical evolution of their host galaxies. We contrasted these results with those obtained from our rare core-collapse supernovae models and compared our models to observations. The main conclusions are as follows.

\begin{enumerate}
  
\item In our fiducial neutron star merger model with natal kicks included, the normalization of the stellar r-process abundances is only slightly lower at solar metallicity than for a model without kicks, even though many neutron star mergers occur outside the ISM of the galaxy. This indicates that the elements produced by these events are not lost, but mix with the inflowing gas that goes on to form stars enriched in r-process elements. Larger differences are seen for lower metallicity stars ($\mathrm{[Mg/H]}\lesssim-1$) where the median $\mathrm{[\rp/Mg]}$ is somewhat lower when kicks are included (even after scaling out the normalization difference at solar metallicity). However, in our `optimized' model with relatively prompt neutron star mergers (using a steeper delay time distribution and shorter minimum delay time), the median r-process abundances are virtually unchanged by including or excluding natal kicks. 

\item The addition of natal kicks strongly decouples the location of supernovae, which occur primarily within the ISM shortly after their progenitor stars were formed, from the location of neutron star mergers, which are most likely to occur outside the ISM. This causes the main effect of kicks: they substantially increase the scatter of  $\mathrm{[\rp/Mg]}$ at all metallicities. This effect becomes larger at lower metallicity. Strongly r-process enhanced stars are far more numerous when kicks are included. In our fiducial model, the percentage of metal-poor stars with $\mathrm{[\rp/Mg]}>1$ increases from 1 to 10 per cent (0.5 to 4 per cent) at $-5<\mathrm{[Mg/H]}<-3$ ($-5<\mathrm{[Mg/H]}<-2$) with the addition of strong natal kicks. 
  
\item The r-process abundance ratios are not affected much by changing details about the kick velocity distribution, as long as the average velocities are large ($v_\mathrm{kick}\gtrsim100$~km~s$^{-1}$, though the exact value cannot be determined accurately with our current set of models). Smaller kick velocities ($v_\mathrm{kick}\lesssim100$~km~s$^{-1}$) indeed result in decreased scatter in $\mathrm{[\rp/Mg]}$. However, even a model with relatively prompt neutron star mergers (using a steeper delay time distribution and shorter minimum delay time) and with a small natal kicks of $v_\mathrm{kick}=0\pm50$~km~s$^{-1}$ still resulted in an increased scatter compared to a model without kicks. We therefore conclude that natal kicks are likely important and need to be included in galactic chemical modelling efforts. 

\item A special, rare type of core-collapse supernova that produces r-process elements results in a relatively flat median $\mathrm{[\rp/Mg]}$ as a function of metallicity, although the median does drop rapidly towards the lowest metallicities at $\mathrm{[Mg/H]}<-4$ ($\mathrm{[Mg/H]}<-3$) for a model with 1 in 1,000 (1 in 10,000) supernovae producing r-process elements. The scatter tends to be lower than in our neutron star merger models and there are fewer extreme r-process enhanced outliers. 

\item As expected based on our previous work \citep{Voort2020}, our fiducial model does not match currently available observations of stellar r-process abundances in detail. These observations suggest a fairly flat relation between the median $\mathrm{[\rp/Mg]}$ and metallicity, which is something quite naturally achieved when rare core-collapse supernovae are responsible for producing r-process elements. However, in our fiducial neutron star merger model, we find a rising trend with metallicity. We therefore tweaked (or `optimized') the parameters of the delay time distribution, within the allowed range, and increased the number of neutron star mergers at early times and thus at low metallicity. Our simulations show that this can also reproduce the same flat trend as seen in observations.

\item Interestingly, the model with `optimized' delay time distribution parameters, but without natal kicks, does not reproduce the observed scatter in $\mathrm{[\rp/Mg]}$. However, the same model where natal kicks were included matches the observed scatter very well. If this `optimized' delay time distribution turns out to be an accurate representation of the neutron star merger rate in the Universe, then our simulations predict that the scatter in r-process abundances at low metallicity is not caused by the rarity of the r-process producing source, but rather by the shape of its natal kick velocity distribution. 
  
\end{enumerate}

Neutron star kicks and their effect on r-process enrichment are not straightforward to include in more idealized models and additional assumptions need to be made. \citet{Oirschot2019} built a semi-analytic model on top of dark matter-only simulations, using a merger `delay time distribution similar to that in our `short delay' model and a bimodal kick velocity distribution. They found a lower normalization of $\mathrm{[\rp/Mg]}$ when kicks were included. This is similar to what we found in our fiducial neutron star merger model, although our `optimized' model did not show this behaviour. Their model did not produce sufficient stars with high r-process abundances at low metallicities, which they attributed to their assumption of instantaneous mixing throughout the galaxy, which meant that the enrichment was too homogeneous. Our simulations track the full inhomogeneous chemical evolution on all scales down to our resolution limit and this likely explains why our models can create a substantial population of r-process enhanced stars.

Though we did not discuss the evolution of iron in this work, it is of note that both $\mathrm{[Mg/Fe]}$ and $\mathrm{[\rp/Fe]}$ are observed to decrease with metallicity at $\mathrm{[Fe/H]}\gtrsim-1$, which has been argued is difficult to reproduce in enrichment models based on neutron star mergers alone \citep{Cote2019}. \citet{Banerjee2020} included inside-out disc evolution and neutron star natal kicks into a one-zone galactic chemical evolution model and found that this resulted in a clear decreasing trend of $\mathrm{[\rp/Fe]}$ with $\mathrm{[Fe/H]}$ at high metallicity, whereas the relation is flat without kicks. In comparison, our kick model has a much smaller impact on the high-metallicity end of the distribution. 

Our new simulations suggest that neutron star mergers could be the only source of r-process elements in the Universe when natal kicks are included. Although our fiducial model does not match observations in detail, we achieve a reasonable match with current observations in our `optimized' model in which the time-dependence of the delay time distribution is steep ($\gamma\approx-1.5$). The included natal kicks improve the agreement with observations by increasing the scatter in r-process abundances. There may be additional ways to reproduce the stellar abundances from observations. For example, a rare type of core-collapse supernovae could potentially also produce the majority of r-process elements, though may struggle to produce sufficient strongly r-process enhanced stars. Reducing the r-process event occurence frequency from 1 in 1,000 core-collapse supernovae to 1 in 10,000 increases the scatter, but decreases the median $\mathrm{[\rp/Mg]}$, resulting in a similar fraction of r-process enhanced stars. Observations of a homogeneous sample of (extremely) metal-poor stars could determine the fraction of r-process enhanced stars and potentially discriminate between neutron star merger models and rare core-collapse supernova models. 

Our neutron star merger model is relatively simple and the parameter values have large uncertainties associated with them. We therefore implemented a variety of models with different parameter values, but were unable to explore all parameter combinations. For example, our merger rates may be lower than estimated from observations \citep{Pol2019}. Additionally, the neutron star merger rate or their resulting r-process yields could also depend on metallicity, which is not included in our current models, because it is not well-constrained. Any future improvements in constraints on the delay time distribution or the kick velocity distribution for neutron star mergers or on the rate of rare r-process producing core-collapse supernovae could be easily tested in our cosmological framework. For example, identifying the host galaxies of neutron star mergers detected via gravitational waves could put strong constraints on their delay time distribution \citep{Safarzadeh2019, Adhikari2020, McCarthy2020}. Future simulations could also improve by capturing more of the multi-phase structure of the ISM instead of using a smooth single-phase ISM model as we did here. This could change the level of mixing within the ISM and therefore potentially increase its chemical inhomogeneity. Here, we have shown that natal kicks should be taken into account when modeling r-process enrichment from neutron star mergers, because they can substantially enhance the scatter in the stellar r-process abundances. 

\section*{Acknowledgements}

We would like to thank Hans-Thomas Janka and Friedrich-Karl Thielemann for interesting discussions and the referee for helpful comments. 
FvdV is supported by a Royal Society University Research Fellowship (URF\textbackslash R1\textbackslash 191703).
RG acknowledges financial support from the Spanish Ministry of Science and Innovation (MICINN) through the Spanish State Research Agency, under the Severo Ochoa Program 2020-2023 (CEX2019-000920-S).
Our fiducial simulation was performed on computing resources provided by the Max Planck Computing and Data Facility in Garching. 
For the simulations used for our resolution tests, the authors gratefully acknowledge the Gauss Centre for Supercomputing e.V.\ (\url{https://www.gauss-centre.eu}) for funding this project (with project code pn68ju) by providing computing time on the GCS Supercomputer SuperMUC-NG at Leibniz Supercomputing Centre (\url{https://www.lrz.de}).
Software used for this work includes NumPy \citep{Harris2020} and Matplotlib \citep{Hunter2007}.

\section*{Data availability}

Data available on request.

\bibliographystyle{mnras}
\bibliography{NSkick}

\bsp

\label{lastpage}

\end{document}